\documentclass[12pt]{iopart}
\usepackage{graphicx}
\usepackage{color}

\newcommand{\jpsi}{J/$\psi$ }
\newcommand{\pT} {$p_{\rm T}$ }

\newcommand{\sNN}{$\sqrt{s_{_{\rm NN}}}$ }
\newcommand{\s}{$\sqrt{s}$ }
\newcommand{\pp}{$p$+$p$ }

\begin{document}

\title[]{Examine the species and beam-energy dependence of particle spectra using Tsallis Statistics}

\author{Ming Shao$^1$, Li Yi$^1$, Zebo Tang$^1$, Hongfang Chen$^1$, Cheng Li$^1$ and Zhangbu Xu$^{1,2}$}
\address{$^1$ Department of Modern Physics, University of Science and Technology of China, 96 Jinzhai Road, Hefei, Anhui 230026, China}
\address{$^2$ Brookhaven National Laboratory, Upton, New York 11973, US}
\ead{zbtang@mail.ustc.edu.cn}
\begin{abstract}
Tsallis Statistics was used to investigate the non-Boltzmann
distribution of particle spectra and their dependence on particle
species and beam energy in the relativistic heavy-ion collisions
at SPS and RHIC. Produced particles are assumed to acquire radial
flow and be of non-extensive statistics at freeze-out. \jpsi and
the particles containing strangeness were examined separately to
study their radial flow and freeze-out.  We found that the strange
hadrons approach equilibrium quickly from peripheral to central
A+A collisions and they tend to decouple earlier from the system
than the light hadrons but with the same final radial flow. These
results provide an alternative picture of freeze-outs: a thermalized
system is produced at partonic phase; the hadronic scattering at
later stage is not enough to maintain the system in equilibrium
and does not increase the radial flow of the copiously produced
light hadrons. The \jpsi in Pb+Pb collisions at SPS is consistent
with early decoupling and obtains little radial flow. The \jpsi
spectra at RHIC are also inconsistent with the bulk flow profile.
\end{abstract}


\maketitle

\section{Introduction}
Identified particle spectra in transverse momenta are one of the
pillars in the major discoveries in high-energy nuclear
physics~\cite{starWhitePaper,phenixWhitePaper,brahmsWhitePaper,phobosWhitePaper}.
It has been demonstrated that the spectral shape is sensitive to
the dynamics of the nucleus-nucleus collisions~\cite{nxu,star_Xi}
and can be used to obtain the radial flow and temperature at
freeze-out. In addition, it has been argued that hadrons
containing strange or charm valence quarks should have smaller
hadronic interaction cross-section, and should decouple from the
system earlier than the hadrons with only light valence
quarks~\cite{starWhitePaper,nxu,star_Xi}. In this scenario, those
strange and charmed hadrons would carry direct information from
the collisions at earlier stage without dilution due to hadronic
scattering at late stage. Particle spectra at SPS and RHIC have
been used to extract this behavior~\cite{nxu,star_Xi}. The
Blast-wave model with Boltzmann-Gibbs (BGBW) distribution has been
applied to SPS and RHIC, and was the basis for the observation of
the early decoupling of multi-strange hadrons~\cite{nxu,star_Xi}.
On the other hand, the evolution with hydrodynamics shows that the
multistrange particle spectra can be well described by the same
hydrodynamics at the same freeze-out as other
hadrons~\cite{Heinz}.

Recently, non-extensive hydrodynamics has been proposed to explain
some essential features in relativistic heavy-ion
collisions~\cite{Tsallis,Biro,tang,Wilk:1999dr,Wilk:2008ue,Wilk:2009nn,De:2007zza,Alberico:1999nh,Osada:2008sw,Biro:2003vz}.
A simplified model with blast-wave assumption and non-extensive
Tsallis statistics for hadrons at freeze-out has been implemented
and applied successfully to describe the RHIC data including all
the multistrange hadrons upto intermediate $p_{\rm
T}$~\cite{tang}. Same approach was extended to fit the hadron
spectra at SPS~\cite{pop}. Although Ref.~\cite{tang} concluded
that a common non-extensive Tsallis Blast-Wave model (TBW) can
describe hadrons with and without strange valence quarks,
Ref.~\cite{pop} claimed that multistrange hadrons and \jpsi freeze
out at much higher temperature  and acquire smaller radial flow
than light hadrons.

In this paper, we perform a systematic study of hadron spectra at
SPS and RHIC top energies with TBW as in \cite{tang}. Detailed
comparisons of spectral shape from model and data are carried out
to assess the degree of this discrepancy. In addition, we
categorize the spectra into all hadrons, strange hadrons,
non-strange hadrons and \jpsi in an attempt to compare their
spectral shapes to the TBW results.

\section{TBW and its fit to RHIC and SPS data}

The same TBW as in Ref.~\cite{tang} is used in the current analyses:
\begin{eqnarray}
\fl \frac{\rmd N}{m_{\rm T} \rmd m_{\rm T}} \propto m_{\rm T}
\int_{-Y}^{+Y}\cosh(y) \rmd y \int_{-\pi}^{+\pi}\rmd\phi \nonumber\\
\times \int_{0}^{R}r\rmd r (1+\frac{q-1}{T}(m_{\rm
T}\cosh(y)\cosh(\rho) \nonumber \\
-p_{\rm T}\sinh(\rho)\cos(\phi)))^{-1/(q-1)} \label{eq1}
\end{eqnarray}

where the left-hand side is invariant differential particle yield
at mid-rapidity, $m_{\rm T}$ and $p_{\rm T}$ are transverse mass
and transverse momentum of the produced particle, $q$ is the
parameter characterizing the degree of non-equilibrium, and $\rho$
is the flow profile growing as $n$-th power ($n=1$) along the
transverse radial direction ($r$) from zero at the center of the
collisions to $\beta_{\rm s}$ at the hard-spherical edge ($R$).

\begin{figure}[bp]
\centering
\includegraphics[width=0.98\textwidth]{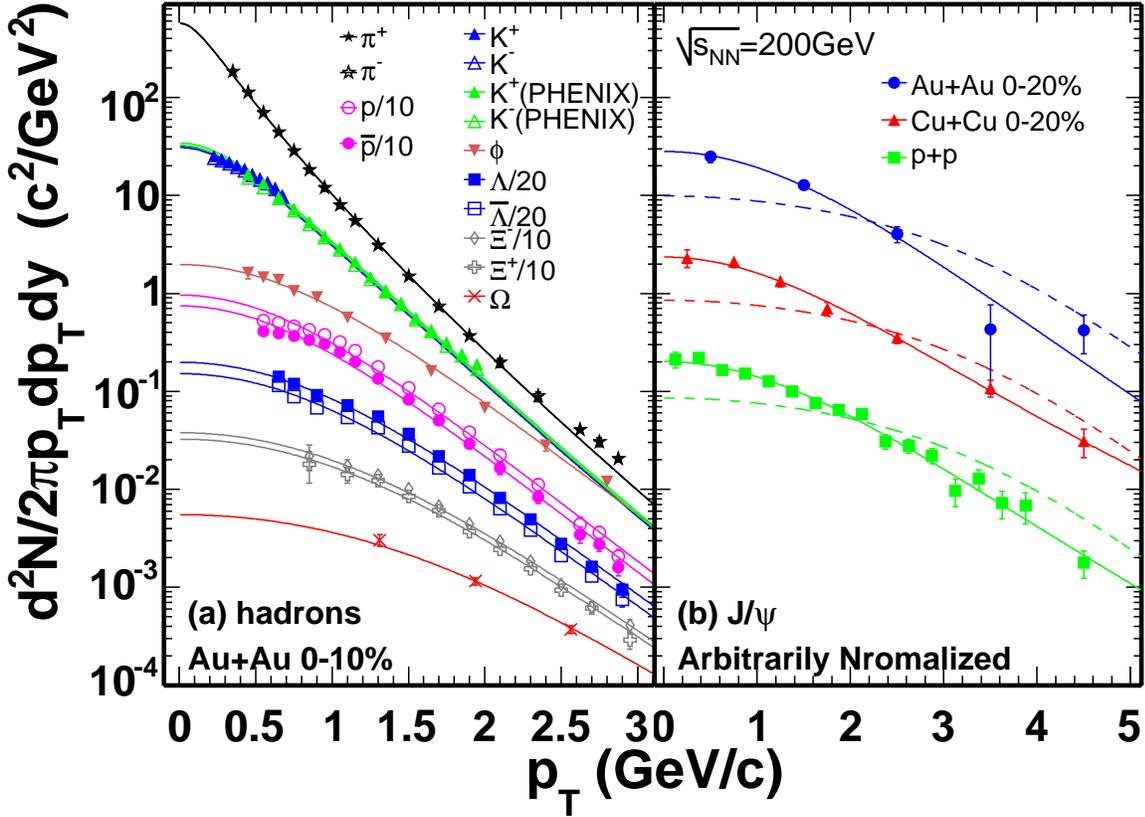}
\caption{(Color Online)
Identified particle spectra from Au+Au collisions at
\sNN=200 GeV. The left panel:  spectra of light hadrons and
strange hadrons. The solid curves are results from TBW fit. the
right panel:  \jpsi spectrum, the dashed line is the TBW
prediction using parameters from the fit to other hadrons, and the
solid curve is a TBW fit to \jpsi alone}\label{fig1}
\end{figure}

The STAR Collaboration has published a series of particle spectra
at mid-rapidity. The most complete set is for p+p and Au+Au
collisions at \sNN= 200 GeV. The identified particle spectra
include $\pi^{\pm}$, $K^{\pm}$, $K_S$, $K^*$, $p$, $\phi$,
$\Lambda$, $\Xi$, $\Omega$, $\bar{p}$, $\bar{\Lambda}$,
$\bar{\Xi}$, and
$\bar{\Omega}$~\cite{starWhitePaper,kstar1,kstar2,ppSplit,
Abelev:2006jr,Abelev:2007ra,Adams:2003qm,Adams:2004ux,
Adams:2006ke,Adams:2006nd}.  In addition, the PHENIX Collaboration
has published spectra of charged kaons~\cite{PHENIXPID} and
$J/\psi$~\cite{PHENIX_Jpsipp,PHENIX_JpsiCuCu,PHENIX_JpsiAuAu},
which are not included in the previous fits~\cite{tang}. The
charged kaon spectra from PHENIX Collaboration extend the STAR
measurement to higher $p_{\rm T}$. This is crucial for the current
study when the spectra are categorized into several groups. In the
group of hadrons with strange valence quarks, kaon is the only
light meson, whose spectrum is more sensitive to the
non-equilibrium effect at low \pT since the spectra of heavier
particles are usually overwhelmed by radial flow in the low-\pT to
intermediate \pT in the central nucleus-nucleus collisions. \jpsi
spectra are measured by PHENIX Collaboration in $p+p$, Cu+Cu and
Au+Au collisions at RHIC top energy for $p_{\rm T}<5$
GeV/$c$~\cite{PHENIX_Jpsipp,PHENIX_JpsiCuCu,PHENIX_JpsiAuAu} while
STAR Collaboration has extended these measurements to high-$p_{\rm
T}$~\cite{STAR_Jpsi}. Only PHENIX data with $p_{\rm T}<5$ GeV/$c$
are used in this TBW analysis. We perform a TBW fit using
\eref{eq1}. The procedure and program are described in details
in~\cite{tang,fuqiang}.

\Fref{fig1} shows the identified spectra measured by STAR and
PHENIX collaborations and their associated TBW curves from the
fit. The yields are presented in terms of invariant differential
cross-section.  There are a few features: the extension of kaon
spectra to high \pT doesn't change the quality of the fit and the
kaon spectra can be well described by the TBW. However, using the
same parameters from the fit, the TBW curve can not reproduce the
\jpsi spectra as shown in the right panel of \fref{fig1}. The
curve is too flat at low $p_{\rm T}$, presumably due to the effect
of large radial flow. We performed a TBW fit to the \jpsi spectra
alone. The result shows that \jpsi has consistently lower radial
flow than the bulk, although the parameter uncertainty is quite
large due to the large statistical uncertainties of the data
points.

\begin{figure}[tbp]
\centering
\includegraphics[width=0.98\textwidth]{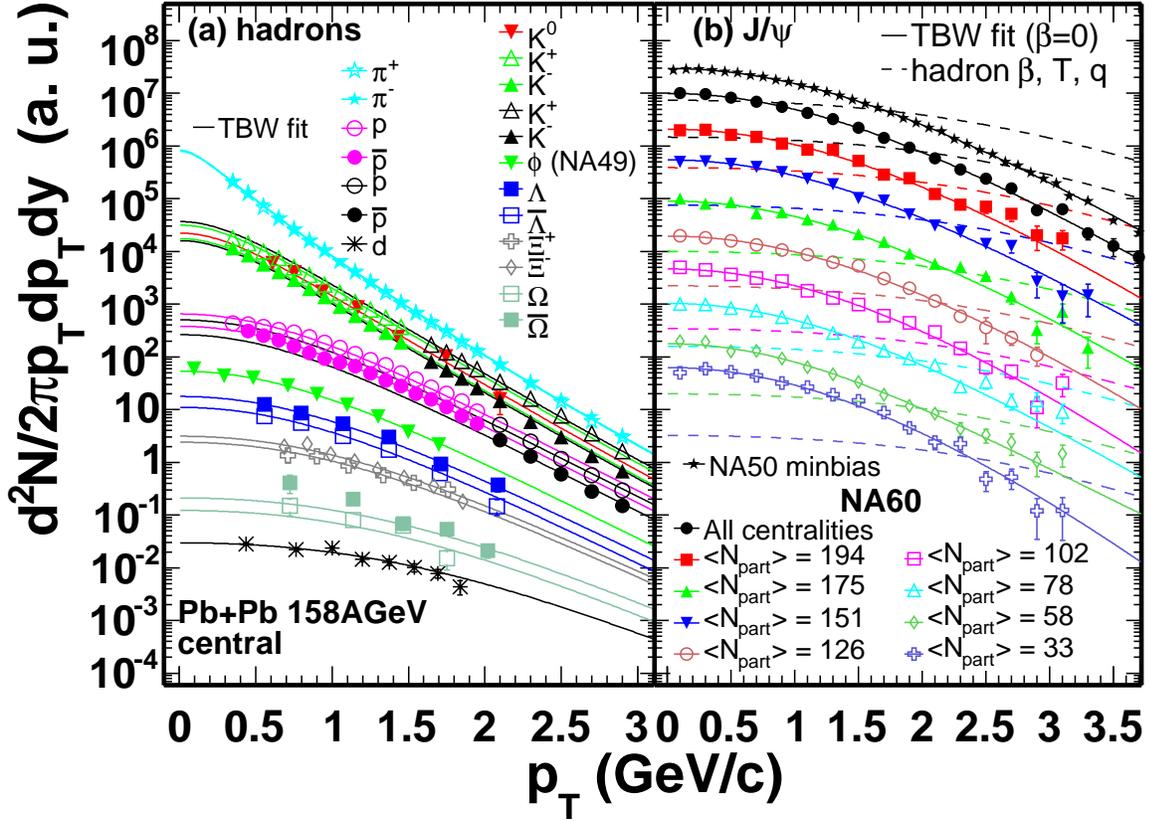}
\caption{(Color Online)
Identified particle spectra in central Pb+Pb collisions
at beam energy of 158 AGeV from the fixed-target experiments at
SPS. The left panel:  spectra of light hadrons and strange
hadrons. The solid curves are results from TBW fit. The right
panel:  \jpsi spectra, the dashed line is the TBW prediction using
parameters from the fit to other hadrons, and the solid curve is a
TBW fit to \jpsi alone.}\label{fig2}
\end{figure}

Similarly, the same TBW fit was applied to SPS
data~\cite{NA49highpt,NA49LambdaXi,NA49Omega,NA49pd,NA49phi,NA50Jpsi,NA57strange,NA57strange2,NA60Jpsi}.
\Fref{fig2} shows the measured spectra together with the fit TBW
curves for central Pb+Pb and In+In collisions at \sNN=17.2 GeV.
The TBW results are generally in good agreement with the data
points. However, the $\chi^2$/nDoF is not as good as the fit at
RHIC. Although the TBW curves represent the trends of the data
points very well, the variations of the data points around the
curves seem to be larger than those allowed by the error bars,
indicating a possible underestimate of the uncertainties. In right
panel of \fref{fig2}, the \jpsi data points from NA50 and NA60
were presented as a function of \pT together with the TBW curves.
The dashed line is the prediction using the parameters produced by
the TBW fit to the other light hadrons, while the solid curve is a
TBW fit to the \jpsi spectrum alone. Same situation as observed at
RHIC, the TBW prediction with parameters obtained from the fit to
the light hadron spectra overpredicts the radial flow present in
the \jpsi spectrum. A fit to \jpsi spectrum alone shows that the
radial flow of \jpsi at SPS is consistent with zero.

\begin{figure}[tbp]
\centering
\includegraphics[width=0.98\textwidth]{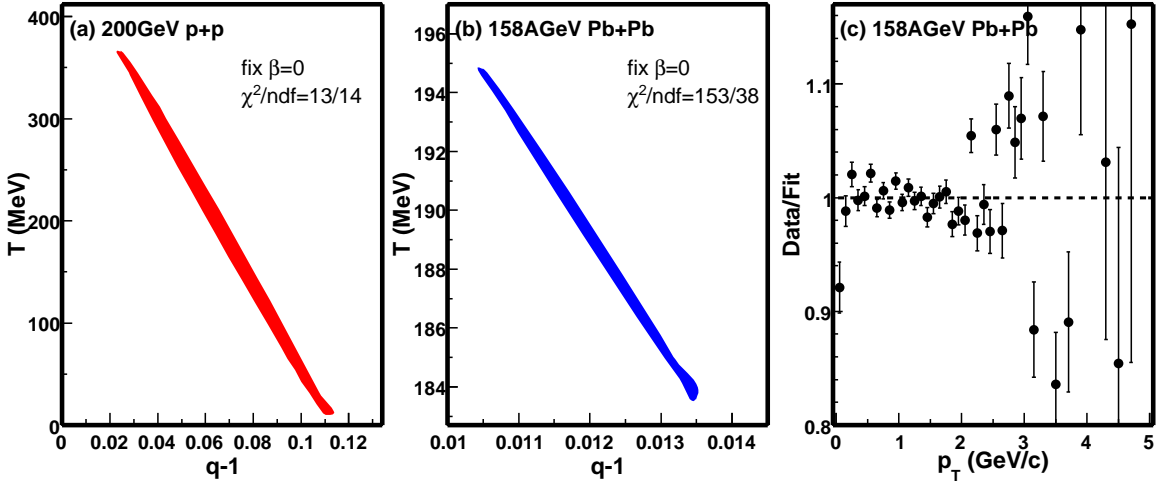}
\caption{ (Color Online)
1-$\sigma$ contour of $T$ vs $(q-1)$ plane from TBW fit
to the \jpsi spectrum in \pp collisions at \s=200 GeV with
$\chi^2$/nDoF=13/14 at minimum (a), Pb+Pb collisions at \sNN=17.2
GeV with $\chi^2$/nDoF=153/38 (b) and ratio of data points and fit
 in Pb+Pb collisions (c). The radial flow is set to zero in this fit.}\label{fig3}
\end{figure}

To further investigate the effect of the correlation between $T$
and $q$ as a potential cause for the large uncertainties from the
\jpsi fit, we examine the contour plot from the TBW fit to the
\jpsi spectra. The \jpsi spectrum in \pp collisions at \s=200 GeV
has much smaller uncertainties with many more data points than
those in A+A collisions at RHIC. We perform a fit by setting
radial flow to zero while only keeping $T$ and $q$ as free
parameters. \Fref{fig3} shows the 1-$\sigma$ contour of the $T$ vs
$(q-1)$ plane with the best fit as presented in \fref{fig1} and
\fref{fig2}. It is obvious from the plot that there is a strong
correlation between these two parameters.  Once one parameter is
constrained accurately by other means or other hadron spectra, the
other parameter is also well constrained.

\begin{figure}[tbp]
\centering
\includegraphics[width=0.98\textwidth]{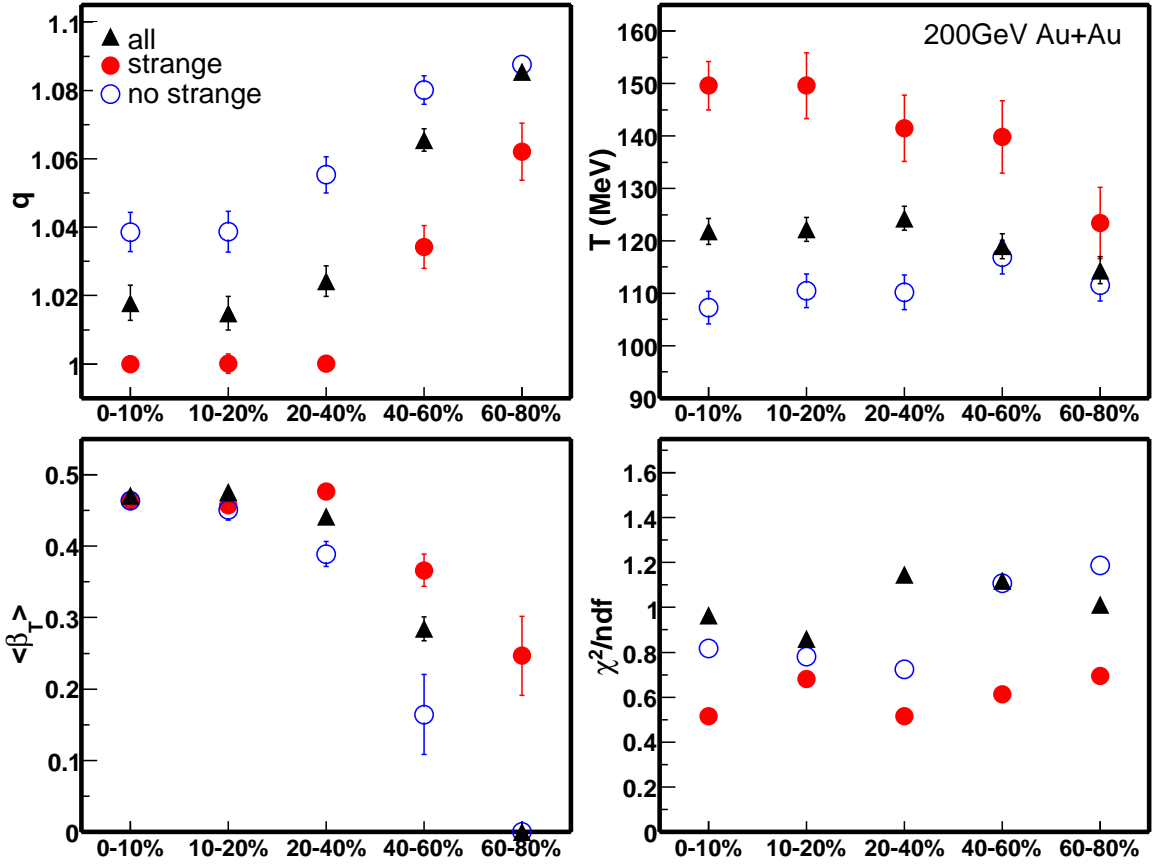}
\caption{(Color Online)
$q$, $T$, $\beta$ and $\chi^2$/nDoF as a function of
centrality for different groups of hadrons from the TBW model fit
to spectra in Au+Au collisions at RHIC top energy.}\label{fig4}
\end{figure}

\Table
{\label{tab1}Fit results for all hadrons, strange hadrons
and non-strange hadrons in central A+A collisions at RHIC and
SPS.} \br
&\centre{3}{Pb+Pb@158AGeV}&\centre{3}{Au+Au@200GeV}\\
\ns
&\crule{3}&\crule{3}\\
&all&strange&non-strange&all&strange&non-strange\\
\mr $\langle \beta \rangle$ (c) & 0.426$\pm$0.004 &
0.420$\pm$0.007 & 0.442$\pm$0.005 & 0.472$\pm$0.009 &
0.464$\pm$0.006 & 0.463$\pm$0.012\\

\mr $T$ (GeV) & 0.113$\pm$0.001 & 0.119$\pm$0.004 &
0.109$\pm$0.001 & 0.122$\pm$0.003 & 0.150$\pm$0.005 & 0.107$\pm$0.004 \\

\mr $q-1$ & 0.015$\pm$0.001 & 0.009$\pm$0.004 & 0.015$\pm$0.001 &
0.017$\pm$0.006 & 0.000$\pm$0.002 & 0.039$\pm$0.006\\

\mr $\chi^2$/nDoF & 267/159 & 137/70 & 102/86 & 140/155 & 51/99 &
43/53\\
\br
\end{tabular}
\end{indented}
\end{table}

Intuitively, one expects that the strange hadrons with smaller
hadronic interaction cross-sections would freeze-out earlier than
the bulk of hadrons~\cite{nxu}. This results in higher temperature
and smaller radial flow for strange hadrons than for light hadrons
without strange content. To study this effect, the particles are
further grouped according to their species: all hadrons, strange
hadrons and non-strange hadrons. \Tref{tab1} shows the fit results
for these groups for central nucleus-nucleus collisions at SPS and
RHIC. Overall, reasonable $\chi^2$/nDoF is present in all cases.
We note that the fits to the strange hadrons produce small
$\chi^2$/nDoF ($\sim$0.5), indicating an experimental overestimate
of the systematic uncertainty. The strange hadrons consistently
show smaller non-equilibrium effect as manifested in a smaller
$(q-1)$ value with higher temperature and similar radial flow than
the non-strange hadrons. On the other hand, the TBW fit to the
strange hadrons at SPS shows larger value of $\chi^2$/nDOF,
indicating less than ideal description of the data. In addition,
the $\Omega$ and $\bar{\Omega}$ data points have larger
fluctuations around the smooth TBW lines than the error bars
allow.  \Fref{fig4} presents the $q$, $T$ and $\beta$ dependence
on the centrality in Au+Au collisions at RHIC top energy. In fact,
the $(q-1)$ is consistent with zero for strange hadron group and
is significantly smaller than that of non-strange hadrons in the
0-40\% central Au+Au collisions at RHIC while they approach each
other in peripheral bins (40-80\%). This is consistent with the
findings of Refs.~\cite{Adams:2006ke,SPS_gammas}: the strangeness
saturation factor $\gamma_s$ increases from peripheral to central
A+A collisions towards unity at RHIC and SPS top energies,
indicating strangeness equilibration in central A+A collisions at
RHIC and SPS.

\section{Discussions}

It is clear that the TBW model can provide satisfactory
description of the spectra in nucleus-nucleus collisions at RHIC
and SPS energies.  There are several important conclusions from
the results:
\begin{enumerate}
\item The radial flow velocity at SPS is smaller than that at
RHIC.

\item Freeze-out temperatures for non-strange hadrons are similar at RHIC and SPS.

\item The non-equilibrium parameter $(q-1)$ is small in central
nucleus-nucleus collisions at RHIC and SPS except a larger $(q-1)$
value for non-strange hadrons at RHIC energy.

\item The categorized groups of strange and non-strange hadrons
show different parameters, indicating a possible early freeze-out
of strange hadrons at RHIC. The biggest difference for strange
hadrons from the light hadrons is the freeze-out temperature:
strange hadrons freeze-out at higher temperature with similar
radial flow. A possible conclusion is that the hadronic phase
doesn't increase the radial flow of light hadrons at RHIC
energies. The picture at SPS is sort of unclear: Although the
strange hadrons have slightly higher temperature, smaller radial
flow, and smaller (q-1) than those of the non-strange hadrons, the
difference is at about $2\sigma$ level. In addition, the $p_T$
reach for the strange hadrons is considerably lower than those for
the non-strange hadrons and the fit quality is less than ideal.

\item The centrality dependence of $(q-1)$ of strange hadrons
shows that the strange hadrons are off-equilibrium in peripheral
collisions as other hadrons. However, they approach thermal
equilibrium very quickly at middle centrality with very small
$(q-1)$ value.

\item Most importantly, the \jpsi spectra at SPS and RHIC require
much smaller flow velocity, and much larger off-equilibrium and/or
higher temperature than the other hadrons. It also shows that the
spectral shapes are the same in \pp and central nucleus-nucleus
collisions. This is inconsistent with thermal production of \jpsi
through recombination of thermal open charms at chemical
freeze-out in the central Au+Au collisions at
RHIC~\cite{PBM,rapp,Thews,Linnyk}. On the other hand, a dynamic
coalescence throughout the system evolution~\cite{Zhuang}
concludes that \jpsi acquires little flow in the process. This
model doesn't maintain \jpsi in equilibrium and is consistent with
our conclusion.
\end{enumerate}

We noted that Ref.~\cite{pop} performed a combined fit to the
strange hadrons and \jpsi, and concluded that both multi-strange
hadrons and \jpsi freeze-out early. However, our separate fits
provide a different picture, where multi-strange hadrons show
deviations from the bulk hadrons with high value of temperature
while \jpsi spectral shape shows little bulk effect, and they are
incompatible with each other.  The forced combination of these two
groups doesn't provide a good fit, but does produce a set of
parameters in-between the separate fit results. We advocate
separate description of strange hadrons and charmed hadrons.
Improved data quality with large kinematic range (especially
$\Omega$ spectra at low $p_{\rm T}$) will provide better fit
results for the multi-strange hadrons. A combination of spectra of
\jpsi and open charmed hadrons and/or their elliptic flow will
enable us to break the correlation among $\beta$, $T$ and $q$. All
these improvements await low-energy scan at RHIC, and future charm
measurements at FAIR, RHIC and
LHC~\cite{HFT_QM09,PHENIX_VXT,LHC_HF,CBM_QM09}.

This study also helps clarify the discrepancy on the claims about
the early freeze-out of multi-strange hadron using a hydrodynamic
evolution and a blast-wave model~\cite{nxu,Heinz}. In the
Boltzmann-Gibbs blast-wave model (BGBW), the obtained $\beta$ for
bulk hadrons ($\pi$,K and p) is about 0.6$c$, much larger than
what we obtained from the TBW model ($\beta\sim0.5c$). This may be
due to the intrinsic assumption of BGBW where the local
thermalization with a Boltzmann distribution increases the radial
flow to absorb the non-equilibrium effect. The large $\beta$
apparently increases the difference to that of multi-strange
hadrons.  On the other hand, it is not clear if freeze-out
temperature is the only variable in the final hydrodynamic
evolution, where the comparison of $\Omega$ spectrum to the
hydrodynamics was made~\cite{Heinz}. Usually, it is argued that a
hybrid model with hadron cascade at the hadronic phase can
describe the experimental data better~\cite{teaney_lauret}. A
least-$\chi^2$ fit to the spectra using different freeze-out
temperature settings in the hydrodynamic model will be valuable at
achieving a quantitatively comparison among the different
approaches. Our fit results in \fref{fig4} show strange hadrons
freeze-out early but their radial flow is the same as that of
light hadrons. This offers an alternative and novel picture of
hadron freeze-out:
\begin{enumerate}

\item At partonic phase, partons approach thermal equilibrium,
resulting in small $(q-1)$ value for strange and light hadrons.
However, \jpsi are not thermalized.

\item After hadronization, multi-strange hadrons decouple from the
system and their measured radial flow reflects that at this stage.

\item At hadronic phase, hadron scattering doesn't produce a
collective radial flow. It also is not sufficient to maintain the
system in equilibrium. The consequence is that the copiously produced light hadrons are much further away from the thermal equilibrium at the end of the hadronic phase than at its beginning.
This results in a large $(q-1$) value for light hadrons
without increasing the radial flow. this may be a consequence
of the explosive nature of the fireball at early stage.

\item This type of investigation is not possible with the BGBW
since equilibrium is built in its assumption.
However, TBW enables us to investigate the effect of non-equilibrium process.

\end{enumerate}

Recently, Tsallis statistics has been applied to extract chemical
freeze-out information at RHIC and SPS~\cite{cleymans,
cleymansSQM09}. However, it shows that the resonance feed-down has
a big impact on the minimum $\chi^2$ of the
fit~\cite{cleymansSQM09}. The chemical fit with Tsallis statistics
doesn't produce a meaningful temperature parameter at the minimum
$\chi^2$ value when the resonance feed-down is applied. It is
tempting to conclude that resonance decays also have a big impact
on the spectral shapes of stable hadrons at the kinetic
freeze-out~\cite{Heinz}. We argue that this is not the case.
Models~\cite{Bleicher, rafelski}, which allow the stable particles
to continue the elastic interactions dominantly through resonances
after chemical freeze-out, can explain the resonance
measurements~\cite{kstar1,kstar2,Lstar} and the extracted
different temperatures between chemical and blast-wave
fits~\cite{starWhitePaper,nxu,star_Xi,tang,fuqiang}. Although the
grand total of each stable particle species including those from
resonance decay at later time is defined to be conserved number
after chemical freeze-out, for very short-lived resonances
produced at chemical freeze-out, most (if not all) of them decay
before the kinetic freeze-out. Meanwhile, their daughters from the
decays interact with the other bulk hadrons in the medium and the
correlation among the daughters as resonance from the chemical
freeze-out has been erased. On the other hand, the stable
particles and resonances after chemical freeze-out continue the
elastic scattering and generate new resonances at a lower rate
before kinetic freeze-out. The resonances available at the kinetic
freeze-out are those regenerated from the bulk hadrons at the time
close to the kinetic freeze-out. Therefore, the resonance yields
and kinematics are completely determined by the kinematics of the
daughter hadrons that generate those resonances, and not the other
way around. As shown in \cite{fuqiang}, the effect of resonance
decay has little impact on the final result. If any, it seems to
support the idea that the resonances (i.e. $\rho$) are from
regeneration with their yields and kinematics determined by the
daughter kinematics.

\section{Conclusions}
Tsallis Statistics in a blast-wave model (TBW) was used to
investigate the non-Boltzmann distribution, radial flow and
freeze-out temperature of particles and their dependence on
particle species and beam energy in the relativistic heavy-ion
collisions at SPS and RHIC. In addition to the light hadrons,
\jpsi and the particles containing strangeness were examined
separately to study their radial flow and freeze-out.  The \jpsi
in Pb+Pb collisions at SPS requires a temperature of 180 MeV and
an average flow velocity of 0.06 $c$ in the TBW fit, and therefore
is consistent with early decoupling and obtains little radial
flow. The \jpsi spectra at RHIC are also inconsistent with the
bulk flow profile. However, there is a strong correlation among
the obtained parameters from the TBW fit to the \jpsi spectrum
along. To improve the uncertainty and break the parameter
correlations, higher statistics, elliptic flow and/or spectra of
other charmed hadrons are needed to further determine the detailed
\jpsi flow pattern at RHIC. We found that the strange hadrons
approach equilibrium quickly from peripheral to central A+A
collisions. The temperature from the TBW fit to strange hadrons is
higher than that for light hadrons while their radial flows are
similar. These results provide a novel picture of freeze-outs: a
thermalized system is produced at partonic phase; the hadronic
scattering at later stage is not enough to maintain the system in
equilibrium and does not increase the radial flow of the copiously
produced light hadrons. The existence of such picture in reality
seems to be natural. However, the investigation of such
off-equilibrium effect has not been possible with the BGBW model
or ideal hydrodynamics.

\ack The authors thank Drs. Lijuan Ruan, Paul Sorensen, Fuqiang
Wang, Pengfei Zhuang, Huanzhong Huang, Gene van Buren, Bedanga
Mohanty, Nu Xu, Jean Cleymans and Jun Takahashi for valuable
discussions and Dr. Enrico Scomparin for providing us with the
NA60 \jpsi data points. This work was supported in part by the
National Natural Science Foundation of China under Grants
10775131, 10805046 and 10835005. Zebo Tang is supported in part by
China Postdoctoral Science Foundation funded project. Zhangbu Xu
is supported in part by the PECASE Grant and by the grant
DE-AC02-98CH10886 from the Offices of NP and HEP within the US DOE
Office of Science.

\section*{References}
\bibliography{Tsallis_SPS_RHIC}

\begin{thebibliography}{10}

\bibitem{starWhitePaper}
John Adams et~al.
\newblock {Experimental and theoretical challenges in the search for the quark
  gluon plasma: The STAR collaboration's critical assessment of the evidence
  from RHIC collisions}.
\newblock {\em Nucl. Phys.}, A757:102--183, 2005.

\bibitem{phenixWhitePaper}
K.~Adcox et~al.
\newblock {Formation of dense partonic matter in relativistic nucleus nucleus
  collisions at RHIC: Experimental evaluation by the PHENIX collaboration}.
\newblock {\em Nucl. Phys.}, A757:184--283, 2005.

\bibitem{brahmsWhitePaper}
I.~Arsene et~al.
\newblock {Quark Gluon Plasma an Color Glass Condensate at RHIC? The
  perspective from the BRAHMS experiment}.
\newblock {\em Nucl. Phys.}, A757:1--27, 2005.

\bibitem{phobosWhitePaper}
B.~B. Back et~al.
\newblock {The PHOBOS perspective on discoveries at RHIC}.
\newblock {\em Nucl. Phys.}, A757:28--101, 2005.

\bibitem{nxu}
H.~van Hecke, H.~Sorge, and N.~Xu.
\newblock {Evidence of early multi-strange hadron freezeout in high energy
  nuclear collisions}.
\newblock {\em Phys. Rev. Lett.}, 81:5764--5767, 1998.

\bibitem{star_Xi}
John Adams et~al.
\newblock {Multi-strange baryon production in Au+Au collisions at \sNN=130
  GeV}.
\newblock {\em Phys. Rev. Lett.}, 92:182301, 2004.

\bibitem{Heinz}
Peter~F. Kolb and Ulrich~W. Heinz.
\newblock {Hydrodynamic description of ultrarelativistic heavy-ion collisions}.
\newblock arXiv:nucl-th/0305084, 2003.

\bibitem{Tsallis}
Constantino Tsallis.
\newblock {Possible Generalization of Boltzmann-Gibbs Statistics}.
\newblock {\em J. Stat. Phys.}, 52:479--487, 1988.

\bibitem{Biro}
Tamas~S. Biro, Gabor Purcsel, and Karoly Urmossy.
\newblock {Non-Extensive Approach to Quark Matter}.
\newblock {\em Eur. Phys. J.}, A40:325--340, 2009.

\bibitem{tang}
Zebo Tang et~al.
\newblock {Spectra and radial flow at RHIC with Tsallis statistics in a
  Blast-Wave description}.
\newblock {\em Phys. Rev.}, C79:051901, 2009.

\bibitem{Wilk:1999dr}
G.~Wilk and Z.~Wlodarczyk.
\newblock {On the interpretation of nonextensive parameter q in Tsallis
  statistics and Levy distributions}.
\newblock {\em Phys. Rev. Lett.}, 84:2770, 2000.

\bibitem{Wilk:2008ue}
Grzegorz Wilk and Zbigniew Wlodarczyk.
\newblock {Power laws in elementary and heavy-ion collisions: A Story of
  fluctuations and nonextensivity?}
\newblock {\em Eur. Phys. J.}, A40:299--312, 2009.

\bibitem{Wilk:2009nn}
Grzegorz Wilk and Zbigniew Wlodarczyk.
\newblock {Multiplicity fluctuations due to the temperature fluctuations in
  high-energy nuclear collisions}.
\newblock {\em Phys. Rev.}, C79:054903, 2009.

\bibitem{De:2007zza}
Bhaskar De, S.~Bhattacharyya, Goutam Sau, and S.~K. Biswas.
\newblock {Non-extensive thermodynamics, heavy ion collisions and particle
  production at RHIC energies}.
\newblock {\em Int. J. Mod. Phys.}, E16:1687--1700, 2007.

\bibitem{Alberico:1999nh}
W.~M. Alberico, A.~Lavagno, and P.~Quarati.
\newblock {Non-extensive statistics, fluctuations and correlations in high
  energy nuclear collisions}.
\newblock {\em Eur. Phys. J.}, C12:499--506, 2000.

\bibitem{Osada:2008sw}
T.~Osada and G.~Wilk.
\newblock {Nonextensive hydrodynamics for relativistic heavy-ion collisions}.
\newblock {\em Phys. Rev.}, C77:044903, 2008.

\bibitem{Biro:2003vz}
Tamas~S. Biro and Berndt Muller.
\newblock {Almost exponential transverse spectra from power law spectra}.
\newblock {\em Phys. Lett.}, B578:78--84, 2004.

\bibitem{pop}
M.~Petrovici and A.~Pop.
\newblock {Collective Phenomena in Heavy Ion Collisions}.
\newblock arXiv:0904.3666, 2009.

\bibitem{kstar1}
C.~Adler et~al.
\newblock {K*(892)$^0$ production in relativistic heavy ion collisions at \sNN
  =130 GeV}.
\newblock {\em Phys. Rev.}, C66:061901, 2002.

\bibitem{kstar2}
John Adams et~al.
\newblock {K*(892) resonance production in Au+Au and \pp collisions at \sNN=200
  GeV at STAR}.
\newblock {\em Phys. Rev.}, C71:064902, 2005.

\bibitem{ppSplit}
B.~I. Abelev et~al.
\newblock {Strange particle production in \pp collisions at \sNN = 200 GeV}.
\newblock {\em Phys. Rev.}, C75:064901, 2007.

\bibitem{Abelev:2006jr}
B.~I. Abelev et~al.
\newblock {Identified baryon and meson distributions at large transverse
  momenta from Au+Au collisions at \sNN=200 GeV}.
\newblock {\em Phys. Rev. Lett.}, 97:152301, 2006.

\bibitem{Abelev:2007ra}
B.~I. Abelev et~al.
\newblock {Energy dependence of $\pi^{\pm}$, $p$ and $\bar{p}$ transverse
  momentum spectrafor Au+Au collisions at $\sqrt{s_{\mathrm {NN}}}$~=~62.4 and
  200 GeV}.
\newblock {\em Phys. Lett.}, B655:104--113, 2007.

\bibitem{Adams:2003qm}
John Adams et~al.
\newblock {Pion, kaon, proton and anti-proton transverse momentum distributions
  from \pp and $d$+Au collisions at \sNN=200 GeV}.
\newblock {\em Phys. Lett.}, B616:8--16, 2005.

\bibitem{Adams:2004ux}
John Adams et~al.
\newblock {$\phi$ meson production in Au+Au and \pp collisions at \sNN=200
  GeV}.
\newblock {\em Phys. Lett.}, B612:181--189, 2005.

\bibitem{Adams:2006ke}
J.~Adams et~al.
\newblock {Scaling Properties of Hyperon Production in Au+Au Collisions at \sNN
  = 200 GeV}.
\newblock {\em Phys. Rev. Lett.}, 98:062301, 2007.

\bibitem{Adams:2006nd}
John Adams et~al.
\newblock {Identified hadron spectra at large transverse momentum in \pp and
  $d$+Au collisions at \sNN=200 GeV}.
\newblock {\em Phys. Lett.}, B637:161--169, 2006.

\bibitem{PHENIXPID}
Stephen~Scott Adler et~al.
\newblock {Identified charged particle spectra and yields in Au+Au collisions
  at \sNN=200 GeV}.
\newblock {\em Phys. Rev.}, C69:034909, 2004.

\bibitem{PHENIX_Jpsipp}
A.~Adare et~al.
\newblock {$J/\psi$ production versus transverse momentum and rapidity in \pp
  collisions at $\sqrt{s}$ = 200 GeV}.
\newblock {\em Phys. Rev. Lett.}, 98:232002, 2007.

\bibitem{PHENIX_JpsiCuCu}
A.~Adare et~al.
\newblock {\jpsi Production in sqrt \sNN=200 GeV Cu+Cu Collisions}.
\newblock {\em Phys. Rev. Lett.}, 101:122301, 2008.

\bibitem{PHENIX_JpsiAuAu}
A.~Adare et~al.
\newblock {\jpsi production vs centrality, transverse momentum, and rapidity in
  Au+Au collisions at \sNN=200 GeV}.
\newblock {\em Phys. Rev. Lett.}, 98:232301, 2007.

\bibitem{STAR_Jpsi}
B.~I. Abelev et~al.
\newblock {\jpsi production at high transverse momentum in \pp and Cu+Cu
  collisions at \sNN=200 GeV}.
\newblock {\em Phys. Rev.}, C80:041902, 2009.

\bibitem{fuqiang}
B.~I. Abelev et~al.
\newblock {Systematic Measurements of Identified Particle Spectra in \pp,
  $d$+Au and Au+Au Collisions from STAR}.
\newblock {\em Phys. Rev.}, C79:034909, 2009.

\bibitem{NA49highpt}
C.~Alt et~al.
\newblock {High Transverse Momentum Hadron Spectra at \sNN=17.3 GeV, in Pb+Pb
  and p+p Collisions, Measured by CERN-NA49}.
\newblock {\em Phys. Rev.}, C77:034906, 2008.

\bibitem{NA49LambdaXi}
C.~Alt et~al.
\newblock {Energy dependence of $\Lambda$ and $\Xi$ production in central Pb+Pb
  collisions at A-20, A-30, A-40, A-80, and A-158 GeV measured at the CERN
  Super Proton Synchrotron}.
\newblock {\em Phys. Rev.}, C78:034918, 2008.

\bibitem{NA49Omega}
C.~Alt et~al.
\newblock {$\Omega^-$ and $\bar{\Omega}^+$ production in central Pb+Pb
  collisions at 40 AGeV and 158 AGeV}.
\newblock {\em Phys. Rev. Lett.}, 94:192301, 2005.

\bibitem{NA49pd}
T.~Anticic et~al.
\newblock {Energy and centrality dependence of deuteron and proton production
  in Pb+Pb collisions at relativistic energies}.
\newblock {\em Phys. Rev.}, C69:024902, 2004.

\bibitem{NA49phi}
C.~Alt et~al.
\newblock {Energy dependence of phi meson production in central Pb+Pb
  collisions at \sNN=6 to 17 GeV}.
\newblock {\em Phys. Rev.}, C78:044907, 2008.

\bibitem{NA50Jpsi}
M.~C. Abreu et~al.
\newblock {Transverse momentum distributions of \jpsi, $\psi'$, Drell- Yan and
  continuum dimuons produced in Pb+Pb interactions at the SPS}.
\newblock {\em Phys. Lett.}, B499:85--96, 2001.

\bibitem{NA57strange}
F.~Antinori et~al.
\newblock {Study of the transverse mass spectra of strange particles in Pb+Pb
  collisions at 158 AGeV}.
\newblock {\em J. Phys.}, G30:823--840, 2004.

\bibitem{NA57strange2}
G.~E. Bruno.
\newblock {Blast-wave analysis of strange particle $m_{\rm T}$ spectra in Pb+Pb
  collisions at the SPS}.
\newblock {\em J. Phys.}, G31:S127--S134, 2005.

\bibitem{NA60Jpsi}
E.~Scomparin.
\newblock {$J/\psi$ production in In+In and $p$+A collisions}.
\newblock {\em J. Phys.}, G34:S463--470, 2007.

\bibitem{SPS_gammas}
F.~Becattini, M.~Gazdzicki, A.~Keranen, J.~Manninen, and R.~Stock.
\newblock {Study of chemical equilibrium in nucleus nucleus collisions at AGS
  and SPS energies}.
\newblock {\em Phys. Rev.}, C69:024905, 2004.

\bibitem{PBM}
A.~Andronic, P.~Braun-Munzinger, K.~Redlich, and J.~Stachel.
\newblock {Statistical hadronization of heavy quarks in ultra- relativistic
  nucleus-nucleus collisions}.
\newblock {\em Nucl. Phys.}, A789:334--356, 2007.

\bibitem{rapp}
L.~Ravagli and R.~Rapp.
\newblock {Quark coalescence based on a transport equation}.
\newblock {\em Phys. Lett.}, B655:126--131, 2007.

\bibitem{Thews}
R.~L. Thews.
\newblock {Quarkonium formation in statistical and kinetic models}.
\newblock {\em Eur. Phys. J.}, C43:97--102, 2005.

\bibitem{Linnyk}
O.~Linnyk, E.~L. Bratkovskaya, and W.~Cassing.
\newblock {Evidence for non-hadronic interactions of charm degrees of freedom
  in heavy-ion collisions at relativistic energies}.
\newblock {\em Nucl. Phys.}, A807:79--104, 2008.

\bibitem{Zhuang}
Li~Yan, Pengfei Zhuang, and Nu~Xu.
\newblock {Competition between \jpsi suppression and regeneration in
  quark-gluon plasma}.
\newblock {\em Phys. Rev. Lett.}, 97:232301, 2006.

\bibitem{HFT_QM09}
Jonathan Bouchet.
\newblock {Heavy Flavor Tracker (HFT) : A new inner tracking device at STAR}.
\newblock arXiv:0907.3407, 2009.

\bibitem{PHENIX_VXT}
H.~van Hecke.
\newblock {Measuring charm and bottom using the PHENIX silicon vertex
  detectors}.
\newblock {\em J. Phys.}, G35:104146, 2008.

\bibitem{LHC_HF}
Andrea Dainese.
\newblock {Charm and beauty at the LHC}.
\newblock {\em Nucl. Phys.}, A783:417--425, 2007.

\bibitem{CBM_QM09}
Johann~M. Heuser.
\newblock {The Compressed Baryonic Matter Experiment at FAIR: Progress with
  feasibility studies and detector developments}.
\newblock arXiv:0907.2136, 2009.

\bibitem{teaney_lauret}
D.~Teaney, J.~Lauret, and E.~V. Shuryak.
\newblock {A hydrodynamic description of heavy ion collisions at the SPS and
  RHIC}.
\newblock arXiv:nucl-th/0110037, 2001.

\bibitem{cleymans}
J.~Cleymans, G.~Hamar, P.~Levai, and S.~Wheaton.
\newblock {Near-thermal equilibrium with Tsallis distributions in heavy ion
  collisions}.
\newblock arXiv:0812.1471, 2008.

\bibitem{cleymansSQM09}
Jean Cleymans.
\newblock Is strangeness chemically equilibrated?
\newblock In {\em Internatianal Conference on Strangeness in Quark Matter
  (SQM)}, 2009.

\bibitem{Bleicher}
Marcus Bleicher.
\newblock {Probing hadronization and freeze-out with multiple strange hadrons
  and strange resonances}.
\newblock {\em Nucl. Phys.}, A715:85--94, 2003.

\bibitem{rafelski}
Giorgio Torrieri and Johann Rafelski.
\newblock {Strange hadron resonances as a signature of freeze-out dynamics}.
\newblock {\em Phys. Lett.}, B509:239--245, 2001.

\bibitem{Lstar}
John Adams et~al.
\newblock {Strange baryon resonance production in \sNN=200 GeV \pp and Au+Au
  collisions}.
\newblock {\em Phys. Rev. Lett.}, 97:132301, 2006.

\end{thebibliography}
\bibliographystyle{unsrt}

\end{document}